\documentclass[a4paper,11pt]{article}
\usepackage{pos}
\usepackage{subcaption}
\usepackage{float}
\usepackage{lineno}

\title{ILC as a SUSY discovery and precision instrument}

\author*[a,1]{M.T. N{\'u}{\~n}ez Pardo de Vera}

\affiliation[a]{DESY,\\
  Notkestrasse 85, Hamburg, Germany}
\note{On behalf of the ILD Concept Group.}

\emailAdd{maria-teresa.nunez-pardo-de-vera@desy.de}

\abstract{
  Data from the LHC at 7, 8, and 13\,TeV have so far yielded no evidence for new particles beyond the 125\,GeV Higgs boson,
  in particular, there have been no signs of SUSY. However, the complementary nature of physics with e$^{+}$e$^{-}$ collisions still
  offers many interesting scenarios in which SUSY can be discovered at the ILC. These scenarios take advantage of the
  capability of e$^{+}$e$^{-}$ collisions to observe events with missing four-momentum - a signature not available at hadron colliders,
  where only transverse imbalance is observable. Due to low backgrounds and trigger-less operation, detectors at e$^{+}$e$^{-}$ colliders
  can observe events with much less visible energy than what is possible at hadron colliders. In this contribution, we will
  present detailed simulation studies done with the ILD concept at the ILC. These studies include simulation of the full
  SM background, as well as realistic accelerator conditions. We will show results both on expected discovery and exclusion
  reaches for the most challenging SUSY channels, such as higgsinos or winos at low mass differences. Evaluations of precision
  of model-parameter measurements, in case of discovery, will also be given. We also report on how such measurements can be
  used to put constraints on parts of the sparticle-spectrum beyond direct reach, and to discriminate between different models
  of SUSY breaking at high scales.}

\FullConference{%
  40th International Conference on High Energy physics - ICHEP2020\\
  July 28 - August 6, 2020\\
  Prague, Czech Republic (virtual meeting)
}

\begin{document}

\maketitle

\begin{section}{Introduction}
  A supersymetric extension of the Standard Model could be the solution or give some light
  into the limitations of this model.
  Up to now there is no evidence of SUSY. SUSY exclusion/discovery
  limits coming from searches at hadron colliders are strongly model
  dependent and they do not cover the whole SUSY parameter space, not being able to access scenarios
  that are needed for SUSY to be the solution of the Standard Model problems.
  The complementary nature of physics at electron-positron colliders, like the ILC, offers the possibility
  of studying these scenarios, computing exclusion/discovery limits in a model independent way and covering
  the parameter space up to the kinematic limit reached by the collider.
  One of the scenarios that would be very challenging or even imposible to be accesed by the
  LHC or any other hadron collider is the one required by SUSY as a solution to the
  absence of electroweak naturalness in the SM. This scenario would imply the existence of a cluster of four light Higgsinos,
  $\widetilde{\chi}_1^{\pm}$, $\widetilde{\chi}_1^0$ and $\widetilde{\chi}_2^0$, within a compressed spectrum (10-20\,GeV)
  and masses around 100-300\,GeV. The experimental conditions at the ILC,
  electron-positron collider at $\sqrt{s}=500$\,GeV with energy upgradability,
  make possible the study of this region of the SUSY parameter space, taking profit of its electron (80$\%$) and
  positron (30$\%$) polarised beams, a well defined initial state (4-momentum and spin configuration), a clean and
  reconstructable final state, hermetic detectors (almost 4$\pi$ coverage) and a triggerless operation.
  This paper is based on two papers showing the capability of the ILC for excluding or, respectively, discovering Higgsinos, and
  more general charginos, in the ILC500 benchmark\footnote{$\sqrt{s}$=500\,GeV, a total integrated luminosity of 4 ab$^{-1}$ divided
    in 1.6 ab$^{-1}$ for $P(e^{-},e^{+})=(-80\%,+30\%)$, 1.6 ab$^{-1}$ for $P(e^{-},e^{+})=(+80\%,-30\%)$, 0.4 ab$^{-1}$ for $P(e^{-},e^{+})=(+80\%,+30\%)$ and 0.4 ab$^{-1}$ for $P(e^{-},e^{+})=(-80\%,-30\%)$}~\cite{chargino_searches} and, in case of discovery,
  for measuring the Higgsino properties with sufficient precision for making important predictions not only related to SUSY but also
  to other subjects in the field of Particle Physics or Cosmology~\cite{higgsino_studies}.
\end{section}

\begin{section}{Chargino searches}
  The lightest chargino is one of the prime candidates to be the Next to Lightest SUSY particle, NLSP, and so the first one to be
  discovered. Some previous studies had shown the reach of the ILC for excluding/discovering charginos, but those studies were done
  with model restrictions and at some specific points of the SUSY parameter space. The study presented in this paper performs chargino searches
  in a general way covering a wide range of parameters.
  The first step of the study was to determine the scenario with the lowest chargino production cross sections assuming only the Minimal
  Supersymetric Standard Model, MSSM, and $R$-parity conservation. The cross sections of this scenario were compared to
  limits for chargino observation extrapolated from LEP studies, what can be also considered as the worst case
  since no improvements due to accelerator or detector technologies were taken into account.
  The cross sections were computed for three different cases depending on the chargino mixing, Higgsino-like, Wino-like
  and mixed charginos, and, since the charginos are produced not only in the $s$-channel but also in the $t$-channel
  via sneutrino exchange, for each of the three different cases the cross sections were computed for low
  (around the kinematic limit) and high (about 1\,TeV) sneutrino masses.
  For high sneutrino masses the worst scenario, lowest cross sections, corresponds to the Higgsino-like charginos.
  If the sneutrino masses are decreased to values close to the kinematic limit the $t$-channel also contributes to
  chargino production, interfering destructively with the $s$-channel. This effect is only seen in the Wino-like case,
  due to the weak coupling between Higgsinos and sneutrinos, and produce a decrease of the production
  cross sections, reaching the minimum values for sneutrino masses at the kinematic limit. This scenario corresponds to
  the lowest cross sections.
 
  The lowest cross-section values for high and low sneutrino masses (Higgsino and Wino-like charginos, respectively) were
  compared to limits from LEP extrapolated to 1.6 ab$^{-1}$, integrated luminosity for $P(e^{-},e^{+})=(-80\%,+30\%)$ polarisation\footnote{The inverse polarisation was not considered in this study since the contribution to the
    chargino production is negligible.} in the ILC500 benchmark, taking only into account the relation between the limits and the inverse of the square root of the
  luminosity, i.e., $Limit_{ILC} = Limit_{LEP}\times\sqrt{\mathcal{L}_{LEP}/\mathcal{L}_{ILC}}$. From this comparison exclusion/discovery limits were determined for Higgsino-like and Wino-like cases. Figures
  ~\ref{Higgsino_ilc_lep_lhc} and \ref{Higgsino_wino_ilc_lep} show the exclusion limits as a function of chargino mass and the
  mass difference between the chargino and the LSP.
  Limits from LEP and from LHC, strongly model dependent and not covering the whole parameter space, are also shown.
  For high sneutrino masses, exclusion and discovery limits are at most 5\,GeV apart from each other. This happens in
  the most challenging region, for values of the chargino-LSP mass difference between the pion mass and 3\,GeV, where the soft products of the chargino
  decay were not energetic enough for triggering the detectors and an ISR photon had to be requested in the trigger,
  decreasing the detection efficiency considerably. Even in the particular case with sneutrino masses close to the kinematic
  limit, not considered in the LEP analyis but included in this study for completeness, exclusion and discovery is guaranteed
  up to less than 10\,GeV from the kinematic limit in the regions with mass difference below the pion mass and above 3\,GeV.
  In the region with mass differences between these values exclusion is guaranteed up to a chargino mass of 225\,GeV and
  discovery up to 205\,GeV. This region has to be taken with care since it was not included in the LEP studies and
  low sneutrino and, in general, sfermion masses would introduce another processes that should be taken into account.
  It is important to note that in the current study none of the ILC-specific improvements in the experimental environment
  were taken into account. A full study including polarisation, the triggerless operation, the improved detectors and the
  tiny size of the ILC beam spot is expected to yield even higher sensitivities.

  \begin{figure}[!htb]
  \begin{subfigure}{0.5\textwidth}
    \includegraphics [width=\linewidth]{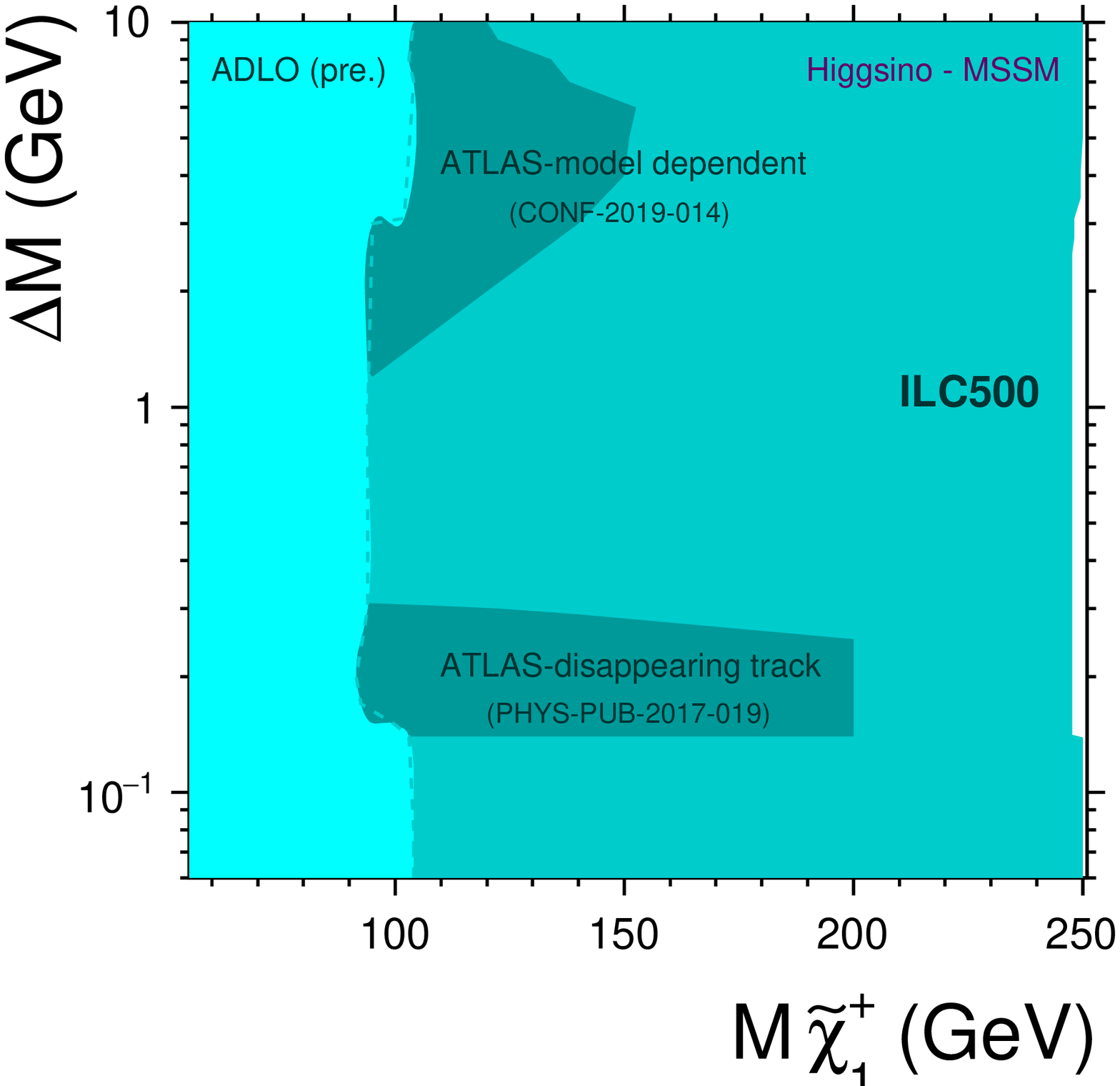}
    \caption{}
    \label{Higgsino_ilc_lep_lhc}
  \end{subfigure} 
  \begin{subfigure}{0.5\textwidth}
    \includegraphics [width=\linewidth]{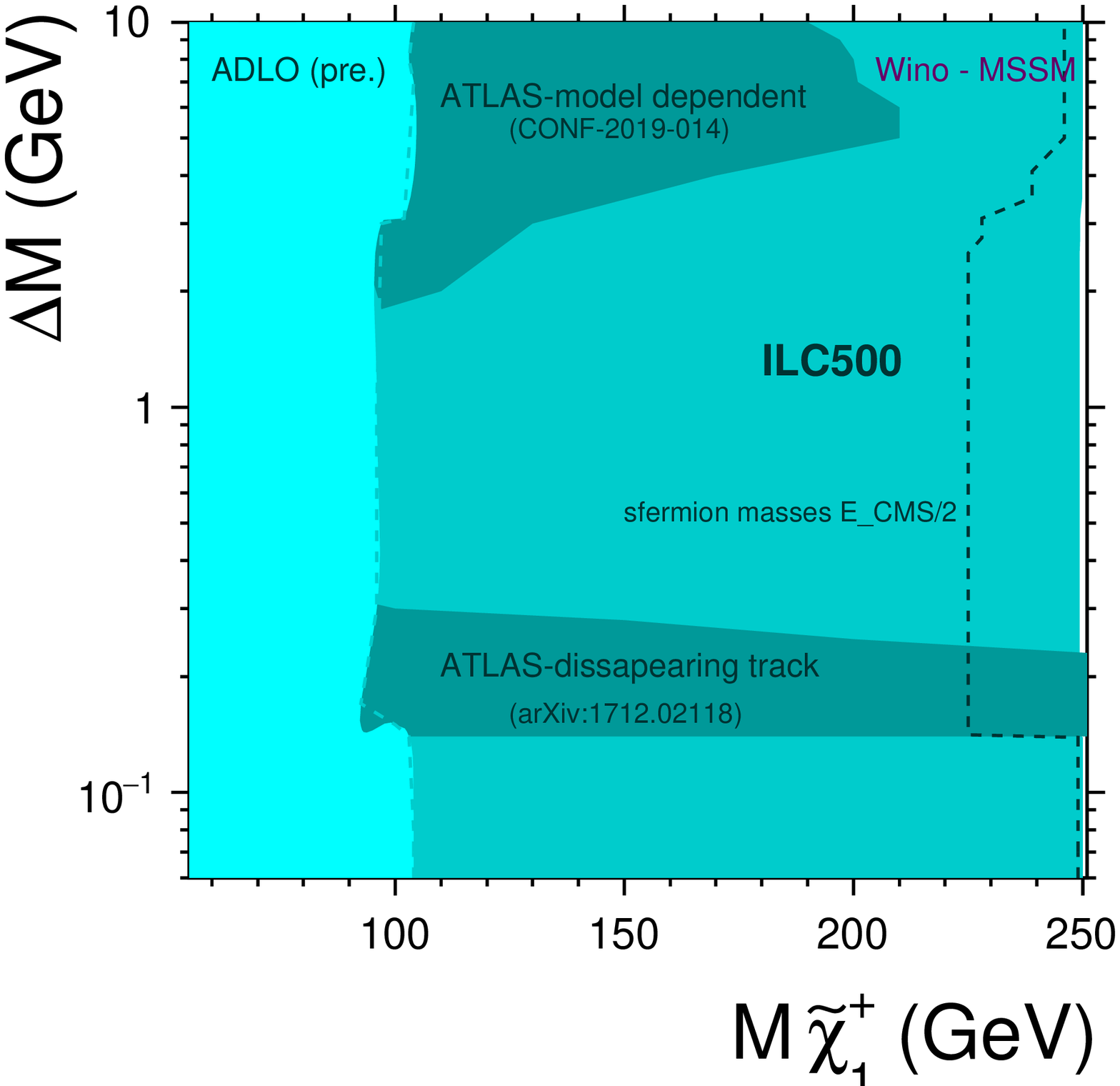}
    \caption{}
    \label{Higgsino_wino_ilc_lep}
  \end{subfigure}
  \caption{(a) Comparison of the $\widetilde{\chi}_1^{\pm}$ mass limits for the Higgsino-like case for LEP, ILC500 and LHC~\cite{ATLAS-Higgsino1}\cite{ATLAS-Higgsino2}. (b) Comparison of the $\widetilde{\chi}_1^{\pm}$ mass limits for the Wino-like case for LEP, ILC500 and LHC~\cite{ATLAS-Higgsino1}\cite{ATLAS-Wino}. The LEP results assume high sfermion masses. In both cases the ATLAS results shown for the region with mass differences above $1$\,GeV are model dependent.}
  \end{figure}

\end{section}

\begin{section}{From light Higgsinos to test of the unification}
  Following a discovery, the ILC has the capability to precisely determine the properties of superparticles.
  Such measurements allow to pin down the underlying model, including the type of SUSY breaking, to predict the
  yet unobserved part of the spectrum and to provide insights into the nature of dark matter.
  All these aspects have been studied for example in three benchmarks with light higgsino-like charginos and
  neutralinos~\cite{higgsino_studies}. The first two ones, ILC1 and ILC2, are based on the NUMH2 model and hence feature
  gaugino mass unification at the GUT scale. The third one, nGMM1, is a mirage mediated
  model with gaugino mass unification at an intermediate scale. These benchmarks were choosen to have
  light Higgsino masses about 150\,GeV and mass differences between 4 and 20\,GeV. The analysis was based on two channels: $e^-e^+\rightarrow \widetilde\chi^+_1\widetilde\chi^-_1\rightarrow \widetilde\chi^0_1\widetilde\chi^0_1qq'\ell\nu_\ell$ and $e^-e^+\rightarrow\widetilde\chi^0_1\widetilde\chi^0_2\rightarrow \widetilde\chi^0_1\widetilde\chi^0_1\ell^+\ell^-$, where $\ell=e$ or $\mu$.
  From kinematic distributions of these channels the masses and mass differences of the Higgsinos, with 1$\%$ precision,
  and their cross-section times the branching ratios, with a few per cent precision depending on the channel and polarisation,
  were extracted.

  \begin{subsection}{Fitting fundamental parameters and testing unification}
    The obtained mass and cross-section projections were used, together with Higgs observables taken from previous ILC studies~\cite{Lehtinen:PhDThesis}~\cite{Yan:2016xyx}, to investigate how the underlying SUSY model can be constrained.

      In a first step, different GUT-scale models were fitted to the projections. Compatibility with the NUMH1 and the CMSSM
      could be ruled out at 95$\%$ CL with only 0.1$\%$ of the total integrated luminosity.
      All three benchmarks were found compatible with the NUMH2 in this approach.

      A more general approach based on weak scale parameters was also investigated. At the tree-level, only four parameters
      enter the Higgsino observables (pMSSM-4). At the one-loop level, ten parameters play a role (pMSSM-10). Already in the
      pMSSM-10 case, all the parameters can be determined with sufficient precision to predict the mass spectrum of the heavier SUSY
      particles, as shown in figure~\ref{fig:ILC1fittedmasses10p}. The masses of
      the heavy electroweakinos and the SUSY higgs bosons can be predicted extremely well, with precisions of about 4$\%$ and
      20$\%$, respectively. The rest of the sfermion masses are less constrained but upper limits can be obtained,
      being very important for future hadron colliders.
      For the pMSSM-4 fit, the loop-level parameters were fixed to the values obtained from the pMSSM-10 fit. As a result
      the predictions for the heavier electroweakino masses improve to 1.6-3$\%$, as shown in figure~\ref{fig:ILC1fitted4p}.
      In addition, it was shown that varying the loop-level parameter values even at the 2$\sigma$ level does not affect
      the pMSSM-4 parameters beyond their 1$\sigma$ range.
      For the other benchmarks analoguous results were obtained.

      \begin{figure}[!htb]
        \begin{subfigure}{0.5\linewidth}
          \includegraphics[width=\textwidth]{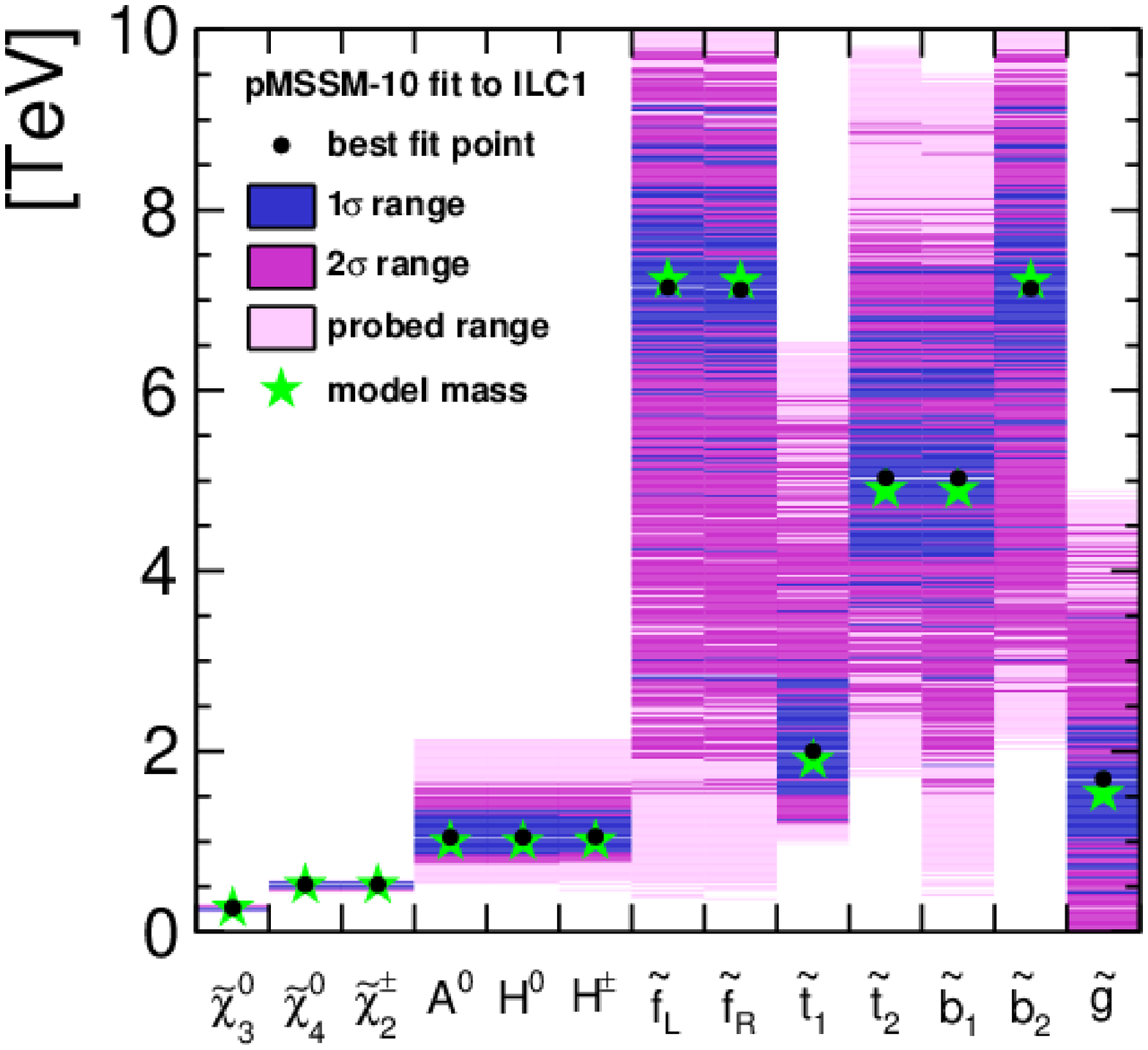}
          \caption{}
          \label{fig:ILC1fittedmasses10p}
        \end{subfigure}
        \begin{subfigure}{0.5\linewidth}
          
          \includegraphics[width=\textwidth]{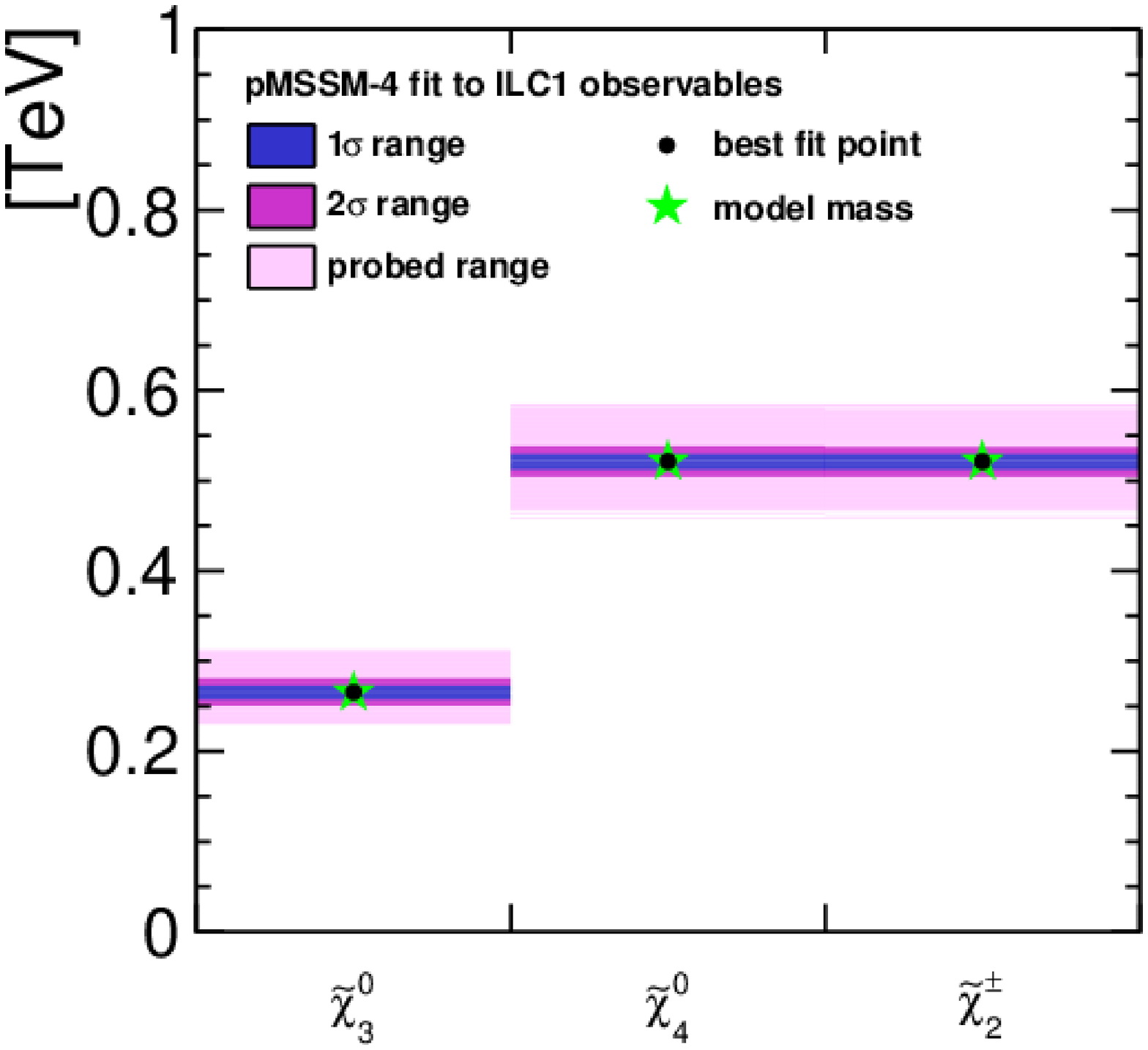}
          \caption{}
          \label{fig:ILC1fitted4p}
        \end{subfigure}
        \caption{(a) Predicted mass ranges from the pMSSM-10 fit to ILC1. (b) Predicted mass and SUSY parameter ranges from the pMSSM-4 fit to ILC1. The green/magenta stars 
            indicate the true model values, while the black dots show the best fit point.}
      \end{figure}

      In a final step, the MSSM renormalization group equations (RGEs) were used to evolve the pMSSM-10 fitted parameters to higher
      energy scales and test the unification hypothesis. Figure~\ref{fig:ILC1-H20-running:M1M2M3} shows the running gaugino masses
      based on the fit to the ILC1 observables, where the bands correspond to 1\,$\sigma$ confidence interval. Running gaugino masses $M_{1}$ and $M_{2}$ cross close
      to 10$^{16}$, verifying the prediction of a SUSY GUT model. The $M_{3}$ band is quite wide but also consistent
      with the unification hypothesis. The unified gaugino mass parameter $M_{1/2}$ is also in agreement with the value obtained from
      the previously mentioned fit of NUMH2 parameters to the observables.
      The same conclusions arise from the fit to the ILC2 observables.
      If the gluino is discovered at the LHC or a future hadron collider, its direct mass determination can be cross-checked against
      the weak scale fit to ILC observables. Alternatively, $M_{3}$ at the weak scale can be predicted from $M_{1/2}$ assuming gaugino
      mass unification.

      The nGMM1 benchmark has the smallest mass differences among the three cases and is thus the experimentally most challenging
      one.
      When assuming precisions of 1.7$\%$ and, in average, 7$\%$ on the masses and cross-sections, respectively, and performing only a pMSSM-10 fit,
      an unification of $M_{1}$ and $M_{2}$ at the GUT scale can be excluded at the 99.9$\%$ CL. However the error bands on the
      running masses are very wide, as shown in figure~\ref{fig:nGMM1-running:M1M2M3}, and it remains unclear whether there is
      an unification at a lower energy scale.
      With the better mass and cross-sections precisions of 0.5$\%$ and  6$\%$ obtained from the ILD full detector simulation study, and when performing the pMSSM-4
      fit after the pMSSM-10 as described above, a much clearer picture shown in~\ref{fig:nGMM1-running:improved} can be obtained.

\begin{figure}[htbp]

  \begin{subfigure}{0.33\linewidth}
    \hspace{-1.9cm}
    \includegraphics[width=1.2\textwidth]{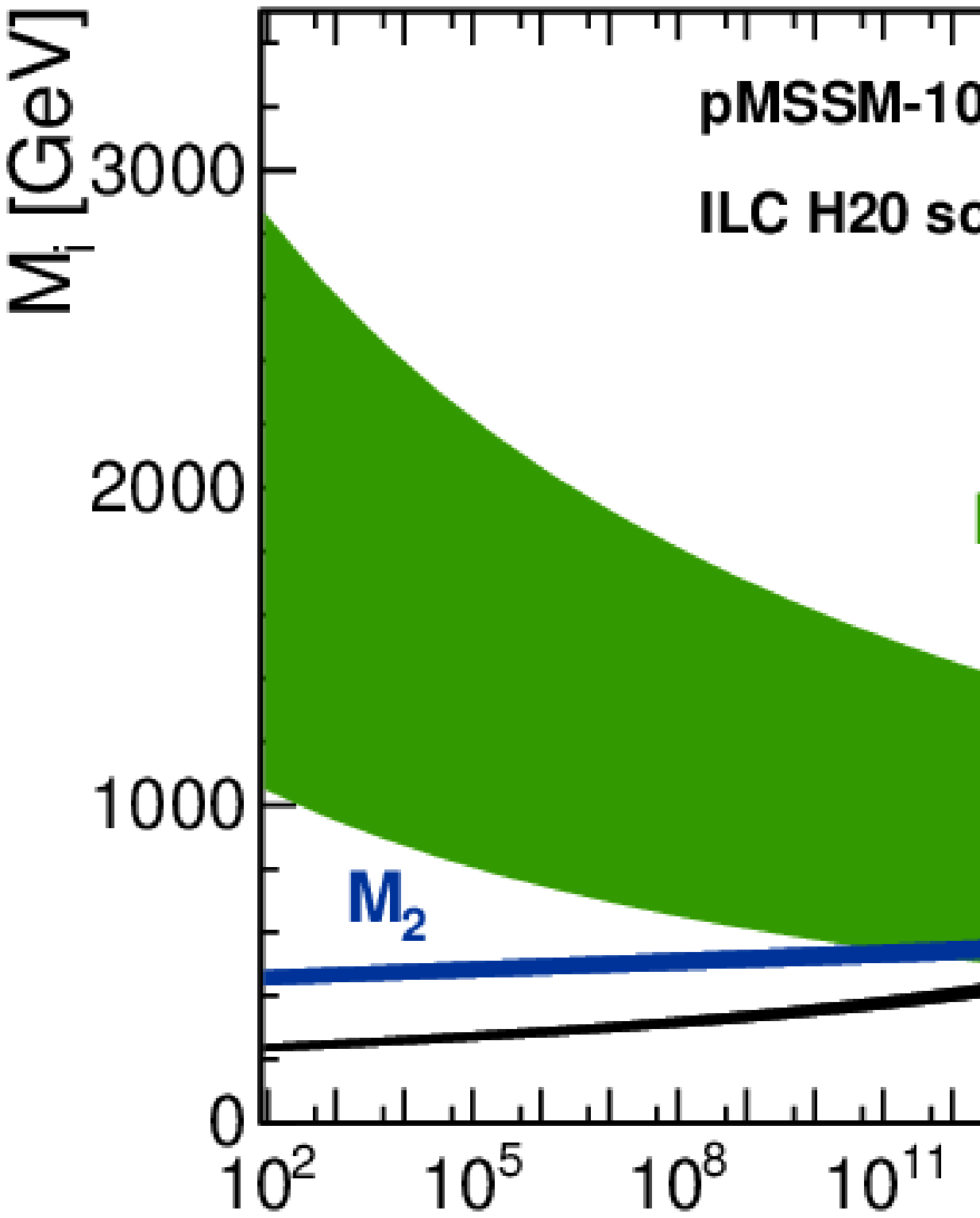}
    \caption{}
    \label{fig:ILC1-H20-running:M1M2M3}
  \end{subfigure}
  \begin{subfigure}{0.33\linewidth}
    \hspace{-1.4cm}
    \includegraphics[width=1.2\textwidth]{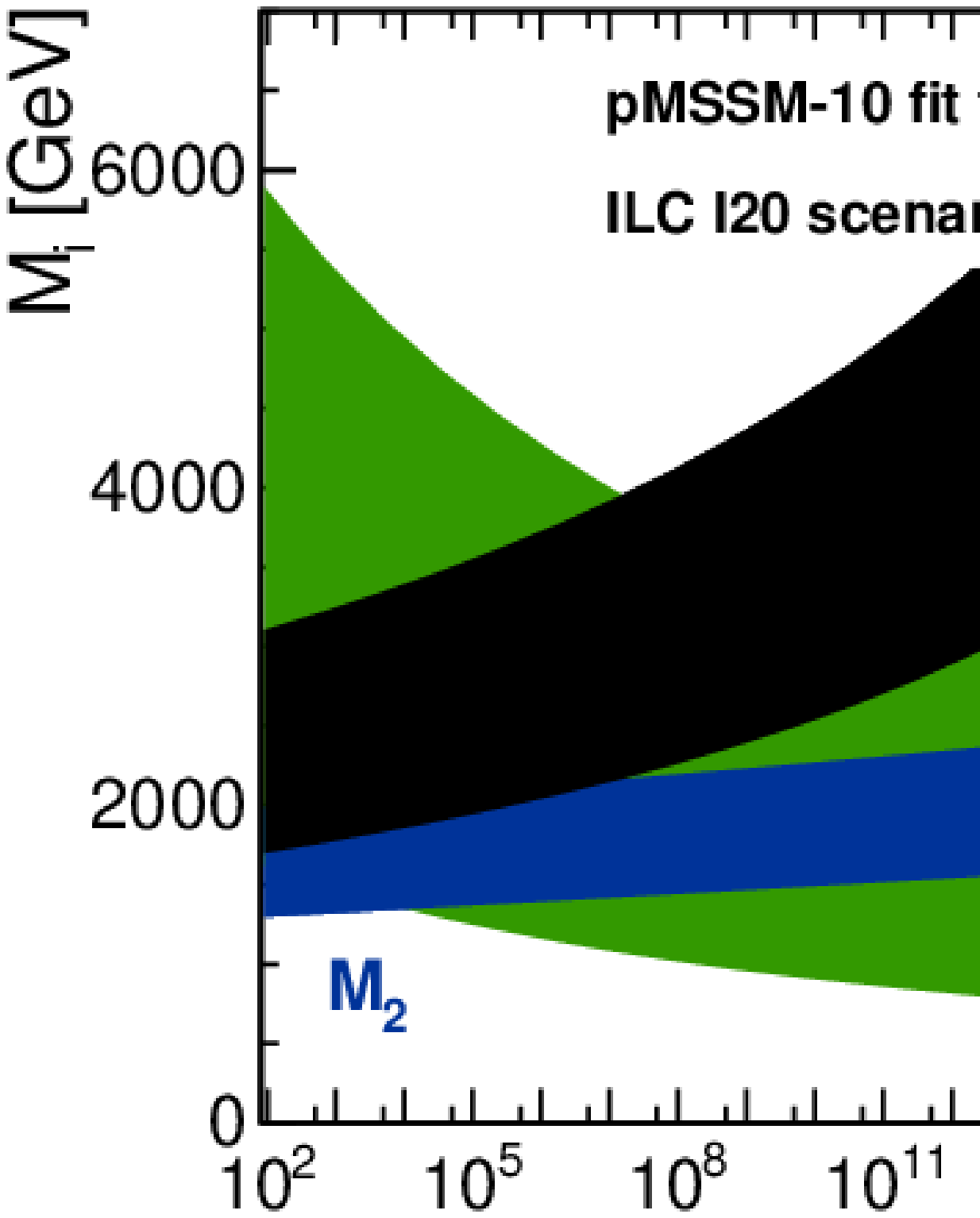}
    \caption{}
    \label{fig:nGMM1-running:M1M2M3}
  \end{subfigure}
  \begin{subfigure}{0.33\linewidth}
    \hspace{-1.05cm}
    \includegraphics[width=1.2\textwidth]{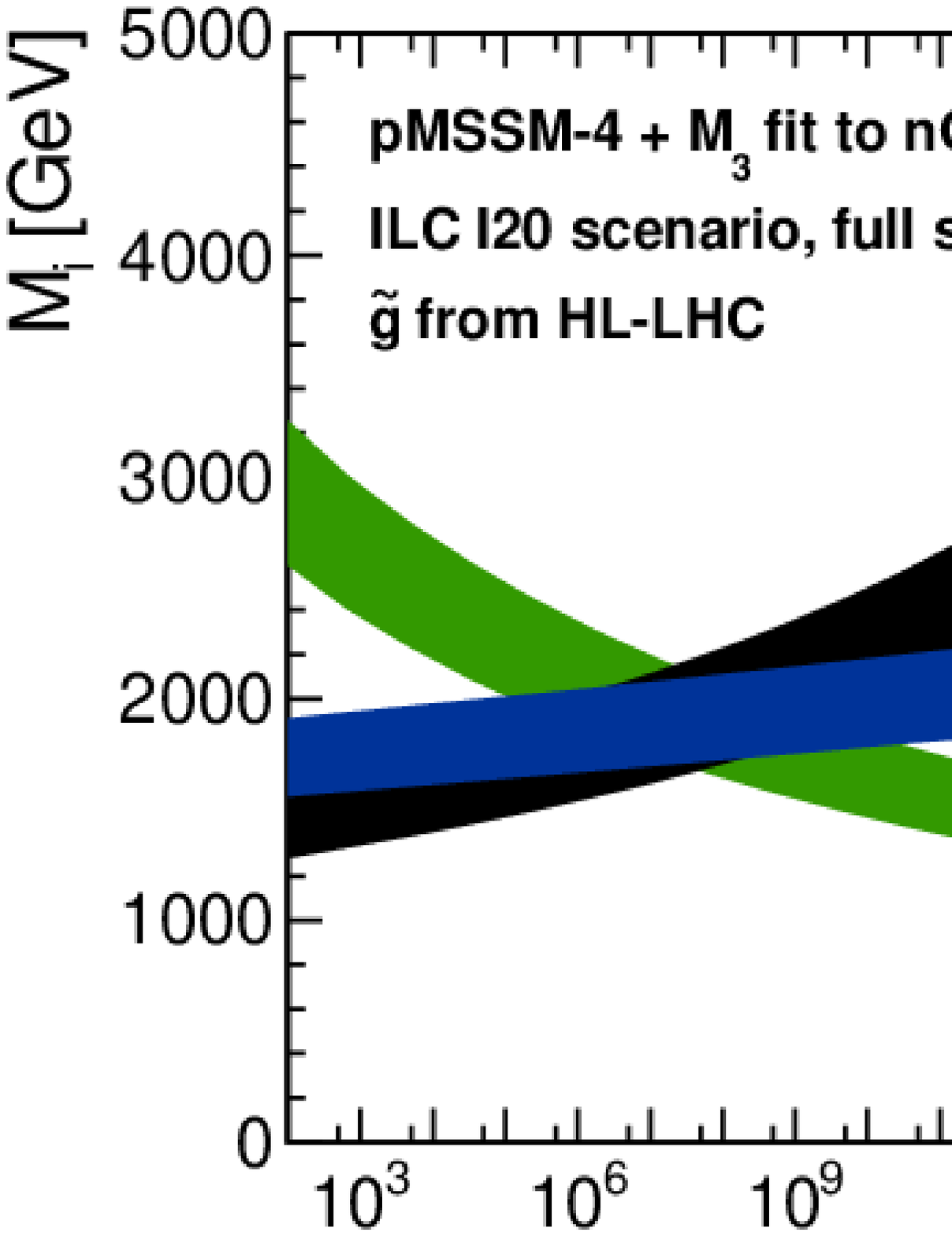}
    \caption{}
    \label{fig:nGMM1-running:improved}
  \end{subfigure}
  \caption{The running gaugino masses $M_i$. Bands correspond to one standard deviation.
    (a) based on the pMSSM-10 fit to ILC1 observables. $M_3$ at the weak scale is used as constrained from ILC measurements.
    (b) and (c) extracting their weak scale values from a fit to nGMM1 observables. (b) pMSSM-10 fit result with
    absolute masses as input. (c) estimated effect of improvement from using the full simulation results, and from including a 10\%
    measurement of the gluino mass from HL-LHC (or other future hadron collider). In addition a fit of the pMSSM-4 parameters, and
    $M_3$ as a fit parameter, is run.}
  \label{fig:nGMM1-running}
\end{figure}
      
  \end{subsection}

\end{section}

\begin{section}{Conclusions}
The reach of the ILC for discovering SUSY in the chargino channel within the worst scenario and in very conservative
conditions has been shown, getting mass limits for exclusion and discovery up to a few GeV below the kinematic limit.
In case of SUSY discovery, the capability of the ILC for measuring SUSY observables with sufficient precision to make
relevant SUSY related predictions has been studied and confirmed, showing how improvements in the experimental results
are cleary reflected in the precision of the predictions.
As a conclusion of the study, it is important to remark how the interplay between ILC and LHC SUSY measurements
or predictions can considerably improve the results of both analyses. 
Moreover, the predictions extracted from direct SUSY measurements can have an important role in accelerator designs
and upgrades beyond the ILC.
\end{section}

\end{document}